\documentclass[12pt, a4paper, fleqn]{article}

\usepackage[nottoc]{tocbibind} 

\RequirePackage[l2tabu, orthodox]{nag}
\usepackage{silence}
\ActivateWarningFilters[pdftoc]

\usepackage[T1]{fontenc}
\usepackage[utf8]{inputenc}

\usepackage[top=20mm, bottom=20mm, left=25mm, right=25mm]{geometry}

\usepackage{array,longtable}
\usepackage{multirow}

\usepackage[table]{xcolor}

\usepackage{amsmath, amssymb, amsfonts, amsthm}
\usepackage{bm}  
\usepackage{mathtools} 

\usepackage{tikz}
\usetikzlibrary{calc,decorations.pathmorphing}
\usetikzlibrary{decorations.pathreplacing}
\usetikzlibrary{shapes}

\usepackage[colorlinks=true, linktoc=all, linkcolor=black, citecolor=red, urlcolor=blue, backref=page]{hyperref}

\usepackage{etoolbox}
\makeatletter
\patchcmd{\BR@backref}{\newblock}{\newblock(}{}{}
\patchcmd{\BR@backref}{\par}{)\par}{}{}
\makeatother

\numberwithin{equation}{section}
\usepackage{pgf}
\usepackage{tikz}
\usetikzlibrary{calc}

\begin{document}

\

\begin{flushleft}
{\bfseries\sffamily\Large 
Fusion rules and structure constants \\ of E-series minimal models
\vspace{1.5cm}
\\
\hrule height .6mm
}
\vspace{1.5cm}

{\bfseries\sffamily 
Rongvoram Nivesvivat$^1$ and 
Sylvain Ribault$^2$
}
\vspace{4mm}

{\textit{\noindent
$^1$
 New York University Abu Dhabi, United Arab Emirates
\\ 
$^2$ Institut de physique théorique, CEA, CNRS, Université Paris-Saclay
}}
\vspace{4mm}

{\textit{E-mail:} \texttt{rongvoram.n@outlook.com,
sylvain.ribault@ipht.fr
}}
\end{flushleft}
\vspace{7mm}

{\noindent\textsc{Abstract:}
In the ADE classification of Virasoro minimal models, the E-series is the sparsest: their central charges $c=1-6\frac{(p-q)^2}{pq}$ are not dense in the half-line $c\in (-\infty,1)$, due to $q=12,18,30$ taking only 3 values --- the Coxeter numbers of $E_6, E_7, E_8$. The E-series is also the least well understood, with few known results beyond the spectrum. 

Here, we use a semi-analytic bootstrap approach for numerically computing 4-point correlation functions. We deduce non-chiral fusion rules, i.e. which 3-point structure constants vanish. 
These vanishings can be explained by constraints from null vectors, interchiral symmetry, simple currents, extended symmetries, permutations, and parity, except in one case for $q=30$.
We conjecture that structure constants are given by a universal expression built from the double Gamma function, times polynomial functions of $\cos(\pi\frac{p}{q})$
with values in $\mathbb{Q}\big(\cos(\frac{\pi}{q})\big)$, which we work out explicitly for $q=12$.

We speculate on generalizing E-series minimal models to generic integer values of $q$, and recovering loop CFTs as $p,q\to \infty$.   
}

\clearpage

\hrule 
\tableofcontents
\vspace{5mm}
\hrule
\vspace{5mm}

\hypersetup{linkcolor=blue}

\section{Introduction}

\subsection{How exceptional are E-series minimal models?}

Virasoro minimal models are the two-dimensional conformal field theories that obey the following assumptions:
\begin{itemize}
 \item \textbf{Rationality}: The spectrum is made of finitely many indecomposable representations of the conformal algebra. 
 \item \textbf{Irreducibility} of all representations that appear.
 \item \textbf{Modular invariance} of the torus partition function. 
\end{itemize}
The CFTs that obey these assumptions can be classified, leading to the ADE classfication of minimal models \cite{cz09}. Minimal models are parametrized by coprime integers $p,q\in\mathbb{N}+2$ that determine their central charges 
\begin{align}
 c_{p,q}=1-6\frac{(p-q)^2}{pq} \ . \label{cpq}
\end{align}
Moreover, these integers obey the following conditions:
\begin{align}
\renewcommand{\arraystretch}{1.3}
 \begin{array}{|c|l|l|}
 \hline 
 \text{Model} & \text{Condition} & \#\{\text{irreducible representations}\}
 \\
 \hline \hline 
  A_{p,q} & q>p & \frac12(p-1)(q-1)
  \\ 
  \hline 
  \multirow{2}{*}{$D_{p,q}$} & q\in 4\mathbb{N}+6 & \frac12(p-1)q
  \\
   & q\in 4\mathbb{N}+8 & \frac12(p-1)(q-1)
  \\
  \hline 
  \multirow{3}{*}{$E_{p,q}$} & q=12 & \frac12(p-1)q 
  \\
   & q=18 & \frac12(p-1)(q-1)
   \\
   & q= 30 & \frac12(p-1)(q+2)
  \\
  \hline 
 \end{array}
 \label{nir}
\end{align}
A-series and D-series minimal models have a prominent place in the landscape of solvable CFTs, thanks to the following features \cite{rib24}:
\begin{itemize}
 \item The spectrums of A-series and D-series minimal models emerge naturally from the representation theory of the Virasoro algebra at rational central charge. 
 \item Their 3-point structure constants are known explicitly, so they can be considered solved. (See \cite{rib24} and references therein.)
 \item The central charges of the A-series and D-series minimal models are dense in the half-line $c\in (-\infty, 1)$. This allows us to take non-rational limits in $c$, and find consistent non-rational CFTs. The converse operation of recovering these minimal models as limits from generic $c$ is also possible to some extent.
 \item Some A-series and D-series minimal models describe critical limits of lattice models that are simple and interesting, such as the Lee--Yang, Ising and 3-state Potts models. 
\end{itemize}
The situation is less favourable for E-series minimal models:
\begin{itemize}
 \item Their spectrums are complicated and strongly $q$-dependent, as we will review in Section \ref{sec:tst}.
 \item Their structure constants are not known explicitly, in spite of some partial results \cite{fgp90}. Structure constants can certainly be computed on a case-by-case basis using standard methods, the problem is that there are many inequivalent cases.
 \item Their central charges are not dense, because $q=12,18,30$ takes finitely many values.
 \item They do not have compelling applications, although some of them describe critical limits of integrable lattice models \cite{pas87}, \cite[Appendix E]{kp07}.
\end{itemize}
We could consider E-series minimal models as eccentric oddities. However, the dearth of exactly solvable CFTs motivates us to entertain another possibility: \textbf{By relaxing rationality, irreducibility and/or modular invariance, could we extend the E-series to infinitely many values of $q$?} 

To extend the E-series, natural candidates would be CFTs whose primary fields belong to the extended Kac table, i.e. their Kac indices would be integer but unbounded. We would expect such CFTs to be irrational and logarithmic. They have sometimes been called logarithmic minimal models, although that expression has never been defined clearly \cite{prz06}, and has also been applied to loop CFTs \cite{rib24} at $c=c_{p,q}$, which have fractional Kac indices. 

Before being able to conclusively answer our question, we however need to better understand the E-series minimal models themselves. In this article we will use bootstrap techniques from loop CFTs, in order to determine fusion rules and structure constants. In the process we will find similarities between loop CFTs and E-series minimal models at the levels of spectrums and structure constants. 

\subsection{The spectrums' tale}\label{sec:tst}

At a given central charge $c_{p,q}$, there may exist up to 3 different minimal models belonging to different series. Consider irreducible, fully degenerate representations of the Virasoro algebra $\mathcal{R}_{r,s}$, parametrized by Kac indices $r,s$, and characterized by their conformal dimensions 
\begin{align}
 \Delta_{r,s} = \frac{(pr-qs)^2 - (p-q)^2}{4pq} \ . \label{drs}
\end{align}
The spectrums of minimal models are built from a set of such representation that is closed under fusion: 
$\left\{ \mathcal{R}_{r,s} \right\}_{(r,s)\in \frac12 ]0,q[ \times ]0, p[ }$, where the factor $\frac12$ accounts for the invariance of $\mathcal{R}_{r,s}$ under $(r,s)\mapsto (q-r,p-s)$, and we introduce the notation 
\begin{align}
 b-a\in\mathbb{N} \quad \implies \quad ]a,b[ = \{a+1, a+2,\dots, b-1\}\ .
\end{align}
The space of states of $A_{p,q}$ is simply the diagonal combination $\frac12 \bigoplus_{r\in ]0,q[}\bigoplus_{s\in ]0,p[} \left|\mathcal{R}_{r,s}\right|^2$, where $\left|\mathcal{R}\right|^2 = \mathcal{R}\otimes \bar{\mathcal{R}}$ is the tensor product of the representation $\mathcal{R}$ of the left-moving Virasoro algebra, with the same representation of the right-moving Virasoro algebra. The space of states of $D_{p, q}$ is not much more complicated, as $D_{p, q}$ can be interpreted as a $\mathbb{Z}_2$ orbifold of $A_{p,q}$. However, the space of states of $E_{p,q}$ is more mysterious, and strongly depends on $q$:
\begin{subequations}\label{specs}
\begin{align}
 \mathcal{S}_{p,12} &=  \bigoplus_{\substack{s\in]0,p[\\ s\in 2\mathbb{Z}+1}}
\left| \mathcal{R}_{1,s}\oplus \mathcal{R}_{7,s}\right|^2
\oplus \frac12 \bigoplus_{s\in]0,p[} 
\left| \mathcal{R}_{4,s} \oplus \mathcal{R}_{8,s}\right|^2
\oplus \bigoplus_{\substack{s\in]0,p[\\ s\in 2\mathbb{Z}+1}}
\left| \mathcal{R}_{5,s} \oplus \mathcal{R}_{11,s} \right|^2 \ ,
\label{s12}
\\
\mathcal{S}_{p,18} &= \frac12 \bigoplus_{s\in]0,p[} \Bigg\{ \left|\mathcal{R}_{9,s}\oplus 2\mathcal{R}_{3,s}\right|^2 \ominus 4\left|\mathcal{R}_{3,s}\right|^2 \oplus \bigoplus_{r\in\{1, 5, 7\}} \left|\mathcal{R}_{r,s}\oplus \mathcal{R}_{18-r,s}\right|^2 \Bigg\} \ ,
\label{s18}
\\
\mathcal{S}_{p,30} &= \frac12 \bigoplus_{s\in]0,p[} \Bigg\{ 
\Bigg|\bigoplus_{r\in\{1,11,19,29\}} \mathcal{R}_{r,s}\Bigg|^2 \oplus
\Bigg|\bigoplus_{r\in\{7,13,17,23\}} \mathcal{R}_{r,s}\Bigg|^2
\Bigg\} \ .
\label{s30}
\end{align}
\end{subequations}
In the cases $q=12$ and $q=30$, the spectrums are block diagonal, and can be decomposed in $3$ and $2$ sectors respectively:
\begin{align}
 \mathcal{S}_{p,12} = \mathcal{S}_{p,12}^{(1)} \oplus \mathcal{S}_{p,12}^{(\sigma)} \oplus \mathcal{S}_{p,12}^{(\epsilon)} \quad , \quad \mathcal{S}_{p,30} = \mathcal{S}_{p,30}^{(1)} \oplus \mathcal{S}_{p,30}^{(\phi)}\ . 
 \label{sectors}
\end{align}
These block-diagonal structures are manifestations of extended symmetries, described by the Ising category for $q =12$ and the Lee--Yang category for $q =30$. Such symmetries also constrain the fusion rules and correlation functions, see Section \ref{sec:es}.

Let us now introduce a more economical notation for the spectrums, which will make the analogy with loop models manifest. So far, we have only used integer Kac indices, and as a result each irreducible representation $\mathcal{R}_{r,s}\otimes \bar{\mathcal{R}}_{\bar r,s}$ depends on 3 indices $(r,\bar r, s)$. However, we may write any pair of left and right dimensions as $(\Delta,\bar \Delta) = (\Delta_{r,s},\Delta_{r,-s})$, provided we allow fractional indices $r,s$ \cite{rib24}. We call $V_{(r,s)}$ the corresponding non-diagonal primary field, whose conformal spin is $\bar\Delta-\Delta =rs\in\mathbb{Z}$. For generic $r,s\in\mathbb{C}$, this primary field would belong to a Verma module of the product of the 2 Virasoro algebras, $V_{(r,s)} \in \mathcal{V}_{\Delta_{r,s}}\otimes \bar{\mathcal{V}}_{\Delta_{r,-s}}$. In our case, the values of the Kac indices such that our primary field belongs to our irreducible representation are
\begin{align}
  V_{\left(\frac{|r-\bar r|}{2},\,  \left(s-\frac{p(r+\bar r)}{2q}\right)\text{sign}(r-\bar r)\right)} \in \mathcal{R}_{r,s}\otimes \bar{\mathcal{R}}_{\bar r,s} \ .
 \label{vrs}
\end{align}
The resulting fields $V_{(r,s)}$ obey $r\in\mathbb{N}^*$ and $rs,qs\in\mathbb{Z}$. For completeness let us also write the reciprocal relation, although it is a bit tricky. Using an integer $a\in\mathbb{Z}$ such that $ap\equiv 1\bmod q$, and writing $\{as\}=as-\lfloor as\rfloor$ the fractional part of $as\in\mathbb{Q}$, we have
\begin{align}
V_{(r,s)} \in \mathcal{R}_{q\, \{as\}-r,p\, \{as\}-s}\otimes \bar{\mathcal{R}}_{q\, \{as\}+r,p\, \{as\}-s} \ .
\end{align}
Using also the notation $V^d_{\langle r,s\rangle}\in \left|\mathcal{R}_{r,s}\right|^2$ for a diagonal field of dimensions $\Delta=\bar\Delta=\Delta_{r,s}$, the spectrums \eqref{specs} are equivalent to the lists of primary fields 
\begin{subequations}
\label{modelsp}
\begin{align}
 \mathcal{P}_{p, 12} &=  \left\{V^d_{\langle r,s\rangle}\right\}_{\substack{r\in\{1,4,7\} \\ s\in]0,p[}}
 \bigcup  \left\{V_{(2,s)}\right\}_{s\in]-\frac{p}{2},\frac{p}{2}[}
 \bigcup \left\{V_{(3,s)}\right\}_{s\in \pm ]-\frac{p}{3},\frac{2p}{3}[}\ ,
 \\
 \mathcal{P}_{p, 18} &= 
  \left\{V^d_{\langle r,s\rangle}\right\}_{\substack{r\in\{1,5,7,9\} \\ s\in]0,p[}} 
 \bigcup \left\{V_{(r,s)}\right\}_{\substack{r\in\{2,4,8\} \\ s\in]-\frac{p}{2},\frac{p}{2}[}}
 \bigcup \left\{V_{(3,s)}\right\}_{s\in \pm ]-\frac{p}{3},\frac{2p}{3}[}\ ,
 \\
 \mathcal{P}_{p, 30} &= 
\left\{V^d_{\langle r,s\rangle}\right\}_{\substack{r\in\{1,7,11,13\} \\ s\in]0,p[}} 
 \bigcup \left\{V_{(2,s)}\right\}_{s\in]-\frac{p}{2},\frac{p}{2}[}
 \bigcup \left\{V_{(r,s)}\right\}_{\substack{r\in\{3,9\} \\ s\in \pm ]-\frac{p}{3},\frac{2p}{3}[}}
 \nonumber \\
 & \hspace{3.7cm}
 \bigcup\left\{V_{(5,s)}\right\}_{s\in \pm ]-\frac{p}{5},\frac{4p}{5}[ }\bigcup\left\{V_{(5,s)}\right\}_{s\in \pm ]-\frac{2p}{5},\frac{3p}{5}[} \ .
\end{align}
\end{subequations}
Let us compare with spectrums of loop CFTs. For definiteness, we write the primary fields of the $PSU(n)$ CFT \cite{rjrs24}:
\begin{align}
 \mathcal{P}_{PSU(n)} = \left\{V^d_{\langle 1,s\rangle}\right\}_{s\in\mathbb{N}^*} \bigcup \left\{V_{(r,s)}\right\}_{\substack{r\in \mathbb{N}^*\\ s\in\frac{1}{r}\mathbb{Z}}}  \ ,
 \label{sun}
\end{align}
The non-diagonal sectors of E-series minimal models are subsets of the non-diagonal sector of the $PSU(n)$ CFT. The main difference is that the second index of $V_{(r,s)}$ is a fraction whose denominator is a prime factor of $q$, and therefore bounded. This suggests that there could be generalizations $\widetilde{E}_{p,q}$ of E-series minimal models to generic values of $q$, whose sets of primary fields $\widetilde{\mathcal{P}}_{p, q}$ would be such that
\begin{align}
 \forall \beta_0^2\in (0,\infty)\ , \quad \lim_{\substack{p,q\to\infty\\ \frac{p}{q}\to \beta_0^2}} \widetilde{\mathcal{P}}_{p, q}  = \mathcal{P}_{PSU(n)}\ .
\end{align}
This condition is fulfilled by sets of primary fields of the type
\begin{align}
 \widetilde{\mathcal{P}}_{p, q} = \left\{V^d_{\langle r,s\rangle}\right\}_{\substack{r\in \{1\}\cup D_q \\ s\in]0,p[}} \bigcup \left\{V_{(r,s)}\right\}_{\substack{r\in N_q \\ s\in S_{p, q, r}}} \ ,
 \label{wppq}
\end{align}
provided the sets $D_q, N_q, S_{p, q, r}$ obey
\begin{align}
\renewcommand{\arraystretch}{1.3}
 \left\{\begin{array}{l} D_q \subset \mathbb{N}^* \\
 \lim_{q\to \infty} \min D_q = \infty \end{array}
\right. \quad
\left\{\begin{array}{l}
        N_q \subset\mathbb{N}^* \\ \lim_{q\to\infty} N_q = \mathbb{N}^*
       \end{array}\right.
\quad
\left\{\begin{array}{l}
       S_{p,q,r} \subset \frac{1}{r}\mathbb{Z} \\ \lim_{p,q\to\infty} S_{p,q,r} = \frac{1}{r}\mathbb{Z}
       \end{array}\right.
\end{align}
Then we could further conjecture that $\lim_{\frac{p}{q}\to \beta_0^2} \widetilde{E}_{p,q}$ is the $PSU(n)$ CFT.

The generalizations $\widetilde{E}_{p,q}$ would have to break rationality, irreducibility and/or modular invariance. In loop models, the fields $V_{(r,s)}$ with $s\in\mathbb{Z}^*$ are logarithmic at generic central charge \cite{nr20}: $\widetilde{E}_{p,q}$ might well include such fields, which would break irreducibility and most probably lead to non-rationality as well. However, we have not been able to find new CFTs of that type. The numerical bootstrap methods of Section \ref{sec:bcf} can be used for testing the validity of conjectured spectrums, by looking for crossing-symmetric 4-point functions. But there are many plausible generalizations \eqref{wppq} of E-series spectrums. We have unsucessfully tested some of these generalizations in the cases $q=9, 15$.

\subsection{Bootstrapping correlation functions}\label{sec:bcf}

Minimal models are the simplest and most well-known exactly solvable CFTs, and are commonly used for testing numerical bootstrap methods (or other techniques or ideas) before applying them to more challenging cases. But in this work we numerically bootstrap E-series minimal models in order to actually solve them. This is not as unreasonable as it seems, because these models have not been solved, in the sense that no explicit formula for 3-point structure constants is known.

The analytic bootstrap method \cite{rib24} studies correlation functions involving the diagonal degenerate fields $V^d_{\langle 2,1\rangle}$ and $V^d_{\langle 1,2\rangle}$, and deduces how structure constants for $V_{(r,s)}$ or $V^d_{\langle r,s\rangle}$ behave under $r\to r+2$ or $s\to s+2$. However, in sharp contrast with the A-series and D-series models, E-series minimal models have spectrums that include $V^d_{\langle 1,2\rangle}$ but not $V^d_{\langle 2,1\rangle}$, and that are correspondingly invariant under $s\to s+2$ (within the interval $s\in]0,p[$) but not $r\to r+2$. So we cannot solve them by this method. 

On the other hand, it is possible to use the same numerical bootstrap methods as in loop CFTs, whose main requirement is an exact knowledge of the spectrum \cite{rib24}. Already in loop CFTs, it is then possible to infer exact results from numerical data. This is done by comparing numerical data with a universal ansatz, assuming that the ansatz only misses a polynomial factor \cite{nrj23}. The same ansatz will work for E-series minimal models. 

Nevertheless, adapting the method to loop models is technically challenging, and we will fully solve the $E_{p,12}$ models but not the $E_{p,18}$ or $E_{p,30}$ models. Let us explain why: 
\begin{itemize}
 \item The crucial parameter for the numerical bootstrap is the size of the spectrum, i.e. the number of primary fields = the number of irreducible representations. In minimal models, this number is finite but not particularly small, as it is of the order $\frac12 pq$, see Table \eqref{nir}. 
 \item Virasoro conformal blocks are numerically computed using Zamolodchikov's recursive representation. In this representation, blocks are written as sums of terms that have poles at rational values of the central charge $c$, even though the blocks themselves are smooth. In loop models, we can take complex values of $c$, far away from all poles. For minimal models, we have to take values of $c$ that differ from the desired rational values \eqref{cpq} by small amounts. The calculations therefore have one more numerical parameter, and one more potential source of systematic errors and numerical instability.  
 \item Inferring exact results from numerical data is a tedious task, done on a case-by-case basis. In the $E_{p,12}$ model we find $13$ nonzero 3-point structure constants that are independent modulo symmetries. This number should scale as $O(q^3)$, and we expect $\sim 44$ structure constants in the $E_{p,18}$ case and $\sim 203$ structure constants in the $E_{p,30}$ case.
\end{itemize}
We have used existing Python code in the public repository \cite{rng24} for bootstrapping $E_{p, 12}$ models. In that repository we have added a notebook for computing structure constants and checking crossing symmetry in $E_{p, 12}$ models. The notebook also computes fusion products in all E-series models.

\subsection{Chiral and non-chiral fusion rules}\label{sec:fr}

In order to interpret our numerical results on 3-point structure constants, we will use the concept of \textbf{non-chiral fusion rules}, which we now define after recalling the more familiar chiral fusion rules:
\begin{itemize}
 \item \textbf{Chiral fusion rules} describe the fusion product of representations of the chiral algebra.
 Given 2 representations $\mathcal{R}_i,\mathcal{R}_j$, we write 
 \begin{align}
  \mathcal{R}_i\times \mathcal{R}_j \simeq \bigoplus_k N_{i,j}^k \mathcal{R}_k  \  \iff \ N_{i,j}^k = \dim\text{Hom}\left(\mathcal{R}_i\times \mathcal{R}_j, \mathcal{R}_k\right)\ . 
 \end{align}
 \item \textbf{Non-chiral fusion rules} describe the fields that appear in operator product expansions. If we write a primary field as $V_I=V_{(i,\bar i)} \in \mathcal{R}_i\otimes \bar{\mathcal{R}}_{\bar i}$, the OPE of 2 primary fields reads 
 \begin{align}
  V_I V_J \sim \sum_K \sum_{\alpha=1}^{M_{I,J}^K} C_{I,J}^{K, (\alpha)} \left(V_K + \text{descendants}_{(\alpha)}\right)\quad \text{with} \quad C_{I,J}^{K, (\alpha)} \neq 0\ ,
  \label{ope}
 \end{align}
 then we define the non-chiral fusion rules as 
 \begin{align}
  V_I* V_J = \sum_K M_{I,J}^K V_K \ . 
  \label{defnc}
 \end{align}
\end{itemize}
We assume that the representations $\mathcal{R}_i$ belong to a category where the fusion product is well-defined and associative. Then associativity implies $\sum_\ell N_{i,j}^\ell N_{\ell,k}^m = \sum_\ell N_{j,k}^\ell N_{i,\ell}^m$. In the case of the Virasoro algebra, fusion multiplicities are trivial, i.e. $N_{i,j}^k \in \{0,1\}$. For W-algebras the fusion multiplicities can be arbitrary integers, or even infinite. The coefficients $N_{i,j}^k$ obey further algebraic properties such as the Verlinde formula.

Chiral fusion constrains OPEs, leading to 
\begin{align}
 M_{I,J}^K \leq N_{i,j}^k N_{\bar i,\bar j}^{\bar k} \ ,
 \label{mlnn}
\end{align}
where we recall $I=(i,\bar i)$. 
Indeed, if we had $N_{i,j}^k\geq 2$, then contributions of descendants would not be determined by primaries, as indicated in Eq. \eqref{ope} by their dependence on $\alpha$. In particular, in E-series minimal models, we have $M_{I,J}^K \in\{0,1\}$.

Non-chiral fusion rules are a notation for indicating which primary fields have nonzero OPE coefficients. While OPEs are associative, the non-chiral fusion rules have no reason to be associative, or to obey nice algebraic properties such as the Verlinde formula. See Eq. \eqref{nonass} for an explicit example of non-associativity.

Nevertheless, non-chiral fusion rules can be associative when they involve \textbf{non-chiral simple currents}: primary fields whose OPE with any other primary field involves only one field. In particular, a simple current that squares to the identity field $V_0$ induces a $\mathbb{Z}_2$ symmetry of the non-chiral fusion rules. Indeed, let $V_s$ be such that $V_s*V_I = V_{\sigma(I)}$ and  
$V_s * V_s = V_0$ with $V_0 * V_I = V_I$. Associativity of fusion products that involve $V_s$ implies that $\sigma$ is an involution, and 
\begin{align}
M_{I,J}^K = M_{\sigma(I),\sigma(J)}^K = M_{I,\sigma(J)}^{\sigma(K)} = M_{\sigma(I),J}^{\sigma(K)}\ .
\label{mijk}
\end{align}
We will find non-trivial simple currents in the $E_{p,18}$ and $E_{p,30}$ minimal models. 

In this work, we determine non-chiral fusion rules as outputs of numerical bootstrap calculations.
The inputs of the calculations are the spectrums \eqref{specs} of E-series minimal models, and the chiral fusion rules i.e. constraints from null vectors. We then check that the results obey the other constraints of Section \ref{sec:con}.
In particular, non-chiral fusion rules reflect the interchiral and extended symmetries of the models.

Once we know non-chiral fusion rules, we can use them as inputs and make bootstrap calculations more efficient, by reducing the number of terms in crossing symmetry equations. We did this for determining the structure constants of the $E_{p,12}$ models. Moreover, any subset of fields that is closed under non-chiral fusion rules defines a submodel: a rational CFT that is consistent on the sphere but not on the torus. Submodels of unitary minimal models have been classified, and in the $E_{p,12}$ case the classification can be rederived from the fusion rules \cite{bcm24}.

Our definition of non-chiral fusion rules has already appeared in the literature, but has not seen much use \cite{rs99}. Quite understandably, non-chiral fusion rules have been written more often in special cases where they are associative. These cases include algebras of simple currents, for example algebras of anyons \cite{fuk24}, or algebras of primary fields in free theories \cite{mor20}. Moreover, in diagonal CFTs, non-chiral fusion rules coincide with chiral fusion rules, and are therefore associative.

\section{Constraints on 3-point structure constants}\label{sec:con}

Let us review the known constraints on 3-point structure constants. In Sections \ref{sec:ncfr} and \ref{sec:sc} we will deduce their detailed consequences in E-series minimal models. 

We use the following notations for 2-point and 3-point structure constants, and for OPE coefficients:
\begin{align}
 \left<V_i V_j\right> \propto \delta_{i,j}B_i \quad , \quad \left<V_iV_jV_k\right> \propto C_{i,j,k} \quad , \quad V_iV_j\sim \sum_k C_{i,j}^k V_k \ . 
 \label{23pt}
\end{align}
These objects obey the relations 
\begin{align}
 B_i = C_{i,i,0} \quad , \quad C_{i,j,k} = C_{i,j}^k B_k \ ,
 \label{bc}
\end{align}
where the index $0$ refers to the identity field $V^d_{\langle 1,1\rangle}$. From 2-point and 3-point structure constants, we can assemble 4-point structure constants, which appear in decompositions of 4-point functions into conformal blocks. For example, the decomposition into $s$-channel blocks $\mathcal{G}^{(s)}_k$ reads
\begin{align}
 \Big<V_1(z)V_2 (0)V_3 (\infty)V_4 (1)\Big> = \sum_k D^{(s)}_k \mathcal{G}^{(s)}_k (z) \quad \text{ with } \quad D_k^{(s)} = \frac{C_{1,2,k} C_{k,3,4}}{B_k}\ ,
 \label{4pt}
\end{align}
where we have used the global conformal invariance to fix $\{z_2, z_3, z_4\} = \{0, \infty, 1\}$.

\subsection{Null vectors}

First, 3-point structure constants are constrained by chiral fusion rules, which follow from conformal symmetry and the existence of null vectors. At a central charge $c=c_{p,q}$, the chiral fusion rules of irreducible, fully degenerate representations of the Virasoro algebra are \cite{fms97}
\begin{align}
 \mathcal{R}_{r_1,s_1}\times \mathcal{R}_{r_2,s_2} = \bigoplus_{r_3\overset{2}{=} |r_1-r_2|+1}^{\min(r_1+r_2,2q-r_1-r_2)-1}\ \bigoplus_{s_3\overset{2}{=} |s_1-s_2|+1}^{\min(s_1+s_2,2p-s_1-s_2)-1} \mathcal{R}_{r_3,s_3}\ , 
\end{align}
where $(r_i,s_i)\in ]0,q[\times ]0,p[$, and the sums run by increments of $2$. We denote $N_{\langle r_1,s_1\rangle \langle r_2,s_2\rangle}^{\langle r_3,s_3\rangle}= N_{\langle r_1,s_1\rangle \langle r_2,s_2\rangle \langle r_3,s_3\rangle}$ the chiral fusion coefficients, which are symmetric under permutations of the 3 representations. Due to the ambiguity $\mathcal{R}_{r,s}=\mathcal{R}_{q-r,p-s}$, there are 2 cases where $N_{\langle r_1,s_1\rangle \langle r_2,s_2\rangle \langle r_3,s_3\rangle} \neq 0$:
\begin{align}
\renewcommand{\arraystretch}{1.5}
\begin{array}{|c|c|}
\hline 
 \text{Odd case} & \text{Even case}
 \\
 \hline 
 \begin{array}{r}  2q-r_1-r_2-r_3\in 1+2{\mathbb{N}} \\
 r_i+r_j-r_k \in 1 + 2{\mathbb{N}} \\
2p-s_1-s_2-s_3\in 1+2{\mathbb{N}} \\
 s_i+s_j-s_k \in 1 + 2{\mathbb{N}} \end{array} 
 & 
 \begin{array}{r}  r_1+r_2+r_3-q\in 1+2{\mathbb{N}} \\
 q+r_k-r_i-r_j\in 1+2{\mathbb{N}}  \\
s_1+s_2+s_3-p\in 1+2{\mathbb{N}} \\
 p+s_k-s_i-s_j\in 1+2{\mathbb{N}} \end{array}
\\
\hline 
 \end{array}
 \label{mmfr}
\end{align}
In E-series minimal models, $q$ is even while $p$ is odd. Therefore, the integers $r_1\pm r_2\pm r_3$ are always odd, while the integers $s_1\pm s_2\pm s_3$ can be even or odd, and we name the 2 cases after its parity. After lifting the ambiguity $\mathcal{R}_{r,s}=\mathcal{R}_{q-r,p-s}$ by choosing one value of $r$ in each pair $\{r,q-r\}$, being even or odd will be a well-defined property of any 3-point coupling. 

For example, in the case $q=12$, we have the following chiral fusion rules:
\begin{align}
 \mathcal{R}_{7,s_1}\times \mathcal{R}_{7,s_2} &= \bigoplus_{s_3\overset{2}{=}|s_1-s_2|+1}^{\min(s_1+s_2,2p-s_1-s_1)-1}\Big(\mathcal{R}_{1,s_3}\oplus \mathcal{R}_{3,s_3}\oplus \mathcal{R}_{5,s_3}\oplus \mathcal{R}_{7,s_3}\oplus \mathcal{R}_{9,s_3}\Big)\ .
 \label{r7r7}
\end{align}
Then $N_{\langle 7,s_1\rangle \langle 7,s_2\rangle \langle 1,s_3\rangle}$ is nonzero only in the odd case. On the other hand, $N_{\langle 7,s_1\rangle \langle 7,s_2\rangle \langle 7,s_3\rangle}$ is nonzero in both odd and even cases, due to the term $\mathcal{R}_{5,s_3} = \mathcal{R}_{7,p-s_3}$, and the situation is the same for $N_{\langle 7,s_1\rangle \langle 7,s_2\rangle \langle 3,s_3\rangle}$. Explicitly, the case $s_1=s_2=1$ reads
\begin{align}
 \mathcal{R}_{7,1}\times \mathcal{R}_{7,1} = \mathcal{R}_{1,1} \oplus \left(\mathcal{R}_{3,1}\oplus \mathcal{R}_{3,p-1}\right) \oplus \left(\mathcal{R}_{7,1}\oplus \mathcal{R}_{7,p-1}\right)\ .
\end{align}
In the 3-dimensional space with coordinates $r_i-\frac{q}{2}$, odd couplings belong to the tetrahedron with vertices $\left\{\frac{q}{2}(\epsilon_1,\epsilon_2,\epsilon_3)\right\}_{\substack{\epsilon_i\in\{\pm 1\}\\ \epsilon_1\epsilon_2\epsilon_3=-1}}$, and even couplings to the tetrahedron with the vertices  $\left\{\frac{q}{2}(\epsilon_1,\epsilon_2,\epsilon_3)\right\}_{\substack{\epsilon_i\in\{\pm 1\}\\ \epsilon_1\epsilon_2\epsilon_3=1}}$. The intersection of the two tetrahedrons is the octahedron with vertices $\left\{(\pm \frac{q}{2},0,0),(0,\pm\frac{q}{2},0),(0,0,\pm\frac{q}{2})\right\}$. 
The intersection of the octant $\{r_i-\frac{q}{2}\leq 0\}$ with the even tetrahedron is included in the odd tetrahedron. In the E-series minimal models, we will choose $r\leq \frac{q}{2}$, except in the case $q=12$ where we use $r=7$ rather than $r=5$.
(This convention is because $V^d_{\langle 7, 1\rangle}$ belongs to the identity sector of the extended symmetry \eqref{sectors}, whereas $V^d_{\langle 5,1\rangle}$ belongs to the $\epsilon$ sector.)
As a result, any allowed even coupling will also come with an allowed odd coupling. In other words, for any $(r_1,r_2,r_3)$, we have either no couplings at all, or odd couplings, or even and odd couplings, with even couplings appearing if $r_1+r_2+r_3>q$.

\subsection{Interchiral symmetry}\label{sec:ic}

In all E-series minimal models, there exists a degenerate diagonal field $V^d_{\langle 1,2\rangle}$. The associativity of OPEs that involve this field implies constraints on structure constants. We may view these constraints as manifestations of a symmetry algebra, called an interchiral algebra \cite{rib24}.

The field $V^d_{\langle 1,2\rangle}$ transforms in the representation $\mathcal{R}_{1,2}\otimes \bar{\mathcal{R}}_{1,2}$. The chiral fusion rules of the Virasoro representation $\mathcal{R}_{1,2}$ are 
\begin{align}
 \mathcal{R}_{1,2}\times \mathcal{R}_{r,s} = \delta_{s>1} \mathcal{R}_{r,s-1} \oplus \delta_{s<p-1} \mathcal{R}_{r,s+1}\ . 
\end{align}
The OPE of $V^d_{\langle 1,2\rangle}$ with another primary field produces all primary fields that are allowed by chiral fusion and
belong to the spectrum. The resulting non-chiral fusion rules are easy enough to write if we parametrize fields by integer left and right Kac indices,
\begin{align}
 V^d_{\langle 1,2\rangle} * V_{(r,s)(\bar r,s)} = \delta_{s>1} V_{(r,s-1)(\bar r,s-1)} + \delta_{s<p-1}  V_{(r,s+1)(\bar r,s+1)}\ , 
\end{align}
including, in the diagonal case $r=\bar r$,
\begin{align}
 V^d_{\langle 1,2\rangle} * V^d_{\langle r,s\rangle} = \delta_{s>1} V^d_{\langle r,s-1\rangle} + \delta_{s<p-1}  V^d_{\langle r,s+1\rangle}\ .
\end{align}
Things get a bit trickier with the notation $V_{(r,s)}$ for non-diagonal fields, because the range of $s$ is no longer $]0,p[$, but some other interval of the same length, say $]s_0,s_0+p[$. Then the non-chiral fusion rules are 
\begin{align}
  V^d_{\langle 1,2\rangle} * V_{(r,s)} = \delta_{s>s_0+1} V_{(r,s-1)} + \delta_{s<s_0+p-1}  V_{(r,s+1)}\ .
\end{align}
This leads to constraints on the behaviour of 3-point structure constants under integer shifts of $s_i$. To be precise, we obtain shift equations that relate all nonzero structure constants of the type $C_{(r_1,s_1)(r_2,s_2)(r_3,s_3)}$ with the same $r_i$, and whose $s_1+s_2+s_3$ differ by even integers. Given $r_1,r_2,r_3$, there are only 2 independent structure constants, which we call $C_{r_1,r_2,r_3}^\text{odd}$ and $C_{r_1,r_2,r_3}^\text{even}$. Here odd/even refer to $s_1+s_2+s_3\bmod 2$ in the notations \eqref{specs} where $s_i$ are integers for both diagonal and non-diagonal fields: this is the same odd/even terminology as in the chiral fusion rules \eqref{mmfr}.
On the other hand, $r_i$ refer to the notations \eqref{modelsp} where both diagonal and non-diagonal fields are characterized by a pair of Kac indices.

In fusion rules,
we may omit the second Kac index, while indicating parity constraints by writing a \textbf{superscript ${}^e$ when both even and odd parities are allowed}, whereas by default only odd couplings appear. In the $E_{p,30}$ models there are cases where only even couplings appear, which we will indicate using a superscript ${}^E$. Such cases occur when odd couplings vanish, even though they are allowed by constraints from null vectors.

\subsection{Simple currents}

We will now construct non-chiral simple currents from chiral simple currents. In principle there might exist non-chiral simple currents that are not built from chiral simple currents, but we will not find any in E-series minimal models.

By a chiral simple current we mean a representation whose fusion product with any other representation yields only 1 representation. In E-series minimal models, these are the representations $\mathcal{R}_{1,1}$ and $\mathcal{R}_{1,p-1}$. Their fusion rules are 
\begin{align}
 \mathcal{R}_{1,1}\times \mathcal{R}_{r,s} = \mathcal{R}_{r,s}  \quad , \quad \mathcal{R}_{1,p-1}\times \mathcal{R}_{r,s} = \mathcal{R}_{r,p-s}\ . 
\end{align}
From these chiral simple currents, we can build the diagonal simple current $V^d_{\langle 1,1\rangle}$, which is trivial. However, we can also build the non-diagonal simple current $V_{(1,1)(1,p-1)}=V_{(1,1)(q-1,1)}=V_{(\frac{q}{2}-1,\frac{p}{2}-1)}$, which is non-trivial if it exists. This field does not exist in the $E_{p,12}$ minimal models, in fact its conformal spin would be half-integer. But it exists in the $E_{p,18}$ and $E_{p,30}$ cases, under the names $V_{(8,\frac{p}{2}-1)}$ and $V_{(14,\frac{p}{2}-1)}$ respectively. 

It is not hard to write the non-chiral fusion rules of the simple current $V_{(\frac{q}{2}-1,\frac{p}{2}-1)}$ with itself and with diagonal fields,
\begin{align}
 V_{(\frac{q}{2}-1,\frac{p}{2}-1)} & * V_{(\frac{q}{2}-1,\frac{p}{2}-1)} \ =\  V^d_{\langle 1,1\rangle} \ ,
 \\ 
 V_{(\frac{q}{2}-1,\frac{p}{2}-1)} & * V^d_{\langle r,s\rangle}\ \underset{r< \frac{q}{2}}{=} \ V_{(\frac{q}{2}-r,\frac{p}{2}-s)}\ , 
 \\
 V_{(\frac{q}{2}-1,\frac{p}{2}-1)} & * V^d_{\langle \frac{q}{2},s\rangle}\ = \ V^d_{\langle \frac{q}{2},s\rangle}\ .
\end{align}
Fusion with non-diagonal fields is a bit trickier to write in general, and we will work it out explicitly in each case. 

In minimal models, all simple currents square to the identity field. Therefore, any nontrivial simple current may be interpreted in terms of an extended $\mathbb{Z}_2$ symmetry. Next, we will discuss extended symmetries that do not come from simple currents.

\subsection{Extended symmetries}\label{sec:es}

The block diagonal structures of the $E_{p,12}$ and $E_{p,30}$ spectrums \eqref{sectors} suggest the existence of extended symmetries, beyond the Virasoro symmetry that is assumed by definition of minimal models. And these extended symmetries constrain non-chiral fusion rules \cite[Exercises 10.14 to 10.16]{fms97}.

In the case of $E_{p,12}$ minimal models, the extended symmetry is described by an Ising-type category. The objects are the blocks $\mathbf{1},\sigma,\epsilon$ that appear in Eq. \eqref{sectors}, and the products are 
\begin{align}
& \mathbf{1}\times \mathbf{1}  = \mathbf{1}\quad , \quad 
 \mathbf{1}\times \sigma  = \sigma\quad , \quad 
 \mathbf{1}\times \epsilon  = \epsilon\quad , \quad 
 \nonumber \\ 
 &\epsilon \times \epsilon = \mathbf{1} \quad ,\quad \sigma\times \sigma = \mathbf{1}+\epsilon \quad , \quad \epsilon \times \sigma = \sigma\ . 
\end{align}
It turns out that almost all these constraints already follow from the chiral fusion rules. The one exception is 
\begin{align}
C^\text{even}_{7,7,7}=0\ . 
\label{777e}
\end{align}
In the spectrum \eqref{s12}, the fields $V^d_{\langle 7,s\rangle}$ with $s$ odd belong to the identity sector of the Ising-type category, whereas $V^d_{\langle 7,s\rangle}$ with $s$ even belongs to the $\epsilon$ sector, under the name $V^d_{\langle 5,p-s\rangle}$. The vanishing of the coupling $\left<\mathbf 1 \mathbf 1\epsilon\right>$ therefore leads to the vanishing of $C^\text{even}_{7,7,7}$.

Similarly, in the case of $E_{p, 30}$, the extended symmetry is described by a Lee--Yang-type category, with the fusion rules 
\begin{align}
 \mathbf{1}\times \mathbf{1}  = \mathbf{1} \quad , \quad \mathbf{1}\times \phi  = \phi \quad , \quad \phi\times \phi = \mathbf{1}+\phi \ . 
\end{align}
Here, it is the vanishing of the coupling $\left<\mathbf{1}\mathbf{1}\phi\right>$ that constrains non-chiral fusion rules. To get constraints that do not already follow from chiral fusion rules, we need fields in the $\mathbf{1}$ sector of the type $V_{(r,s)(\bar r,s)}$ with $r,\bar r\notin \{1,29\}$, i.e. fields that are not built from a left or right chiral simple current. Such fields are $V^d_{\langle 11,s\rangle}$ or $V_{(4,s)}$, but these two cases are related by fusion with the non-chiral simple current. Therefore, modulo the $\mathbb{Z}_2$ simple current symmetry, all constraints boil down to 
\begin{align}
 \forall X \in \phi \ , \quad C_{11,11,X}=0\ .
 \label{1111x}
\end{align}

\subsection{Permutations}

Under odd permutations of the 3 fields, any 3-point structure constant picks a sign $(-1)^{S_1+S_2+S_3}$, where $S_i=\Delta_i-\bar\Delta_i$ is the conformal spin. Therefore,
\begin{align}
 S_Y \in 2\mathbb{Z}+1 \quad \implies \quad \forall X\ , \quad C_{X,X,Y}=0\ .
 \label{cxxy}
\end{align}
In particular, in E-series minimal models, all non-diagonal fields of the type $V_{(2,s)}$ have odd spins, therefore $C_{X,X,2}=0$. Similarly, in $E_{p,30}$ minimal models, we have $C_{X,X,14}=0$.

In the case of a non-diagonal field $V_{(r,s)}$ with $r$ odd, the parity of the spin changes under $s\to s+1$. Therefore, for any field $X$, only one of the two couplings $C_{X,X,r}^\text{even}$ or $C_{X,X,r}^\text{odd}$ can be nonzero. The coupling that is allowed by permutation symmetry may or may not be allowed by chiral fusion rules: we will see examples of both cases.

\subsection{Parity}

Crossing symmetry equations are invariant under complex conjugation of positions, and therefore equivalently under the exchange of left and right conformal dimensions. For a non-diagonal field, this exchange amounts to $V_{(r,s)} \to V_{(r,-s)}$. We will use the compact notation $X\to \bar X$ for this operation, with $X=\bar X$ for diagonal fields.

This does not immediately imply that all structure constants are invariant under flipping the signs of $s$ for all non-diagonal fields. In loop models, there indeed exist parity-odd solutions of crossing symmetry \cite{nrj23}. However, in E-series minimal models, we only find parity-even solutions. We therefore assume that all structure constants are parity-even,
\begin{align}
 C_{X,Y,Z}= C_{\bar X,\bar Y,\bar Z}\ .
\end{align}

\section{Non-chiral fusion rules}\label{sec:ncfr}

Before writing the fusion rules explicitly, let us summarize which constraints apply in $E_{p,q}$ minimal models in each one of the cases $q=12,18,30$. In the following table, we indicate constraints that apply in green, while omitting constraints that are redundant with previous constraints:
\begin{center}
\renewcommand{\arraystretch}{1.3}
 \begin{tabular}{|r|c|c|c|}
  \hline 
  Constraints & $E_{p,12}$ & $E_{p,18}$ & $E_{p,30}$ 
  \\
  \hline \hline 
  Null vectors & \cellcolor{green} & \cellcolor{green} & \cellcolor{green}
  \\
  \hline 
  Interchiral symmetry & \cellcolor{green} & \cellcolor{green} & \cellcolor{green}
  \\
  \hline 
  Simple currents & & \cellcolor{green} & \cellcolor{green}
  \\
  \hline 
  Extended symmetry & \cellcolor{green} & & \cellcolor{green}
  \\
  \hline 
  Permutations & & \cellcolor{green} & \cellcolor{green}
  \\
  \hline 
  Parity & \cellcolor{green} & \cellcolor{green} & \cellcolor{green}
  \\
  \hline \hline 
  Unexplained &  &  & \cellcolor{green}
  \\
  \hline 
 \end{tabular}
\end{center}

\subsection{Case $q=12$}

For each field we give a compact notation $D_k$ (diagonal field) or $N_k$ (non-diagonal field), the full name of the field, the range of values of the second Kac index, and 
the values of the integer left and right Kac indices.
\begin{align}
\renewcommand{\arraystretch}{1.5}
 \begin{array}{|c|c|c|c|c|}
  \hline 
  \text{Notation} & \text{Field} & \text{Values of } s & \text{Left} & \text{Right} 
  \\
  \hline \hline 
 D_1 &   V^d_{\langle 1,s\rangle} & ]0, p[ &  (1, s) & (1, s) 
  \\
  \hline 
  D_4 & V^d_{\langle 4,s\rangle} & ]0, p[ &  (4, s) & (4, s) 
  \\
  \hline 
  D_7 & V^d_{\langle 7,s\rangle} & ]0, p[ &  (7, s) & (7, s) 
  \\
  \hline \hline 
  N_2 & V_{(2, s)} & ]-\frac{p}{2},\frac{p}{2}[ & (4,\frac{p}{2}-s) & (8,\frac{p}{2}-s) 
  \\
  \hline 
  N_3 & V_{(3,s)} &  ]-\frac{2p}{3},\frac{p}{3}[ & (1, \frac{p}{3}-s)  & (7,\frac{p}{3}-s) 
  \\
  \hline 
  \bar N_3 & V_{(3,s)} &  ]-\frac{p}{3},\frac{2p}{3}[ & (7, \frac{p}{3}+s)  & (1,\frac{p}{3}+s) 
  \\
  \hline 
 \end{array}
 \label{spec12}
\end{align}
We now write the fusion rules in our compact notation, which omits the second Kac index. The dependence on that index can be recovered as explained in Section \ref{sec:ic}. In particular, the superscript $^e$ appears when that index takes odd and even values, as opposed to having a fixed parity:
\begin{subequations}
 \begin{align}
 D_1* X &= X\quad (\forall X)\ ,
 \\
  D_4* D_4 &= D_1+{}^eD_7+N_3+\bar N_3\ ,
  \\
  D_4* D_7 &= {}^eD_4+{}^eN_2\ ,
  \\
  D_7* D_7 &= D_1+D_7+N_3+\bar N_3\ ,
  \label{d7d7}
  \\
  D_4* N_2 &= {}^eD_7 + N_3+\bar N_3\ ,
  \\
  D_4* N_3 &= D_4+N_2 \ ,
  \\
  D_7* N_2 &= {}^eD_4+{}^eN_2\ ,
  \\
  D_7* N_3 &= D_7 +\bar N_3\ ,
  \\
  N_2* N_2 &= D_1+{}^eD_7+N_3+\bar N_3\ ,
  \\
  N_2* N_3 &= D_4+N_2 \ ,
  \\
  N_3* N_3 &= D_1 + N_3\ ,
  \\
  N_3* \bar N_3 &= D_7\ .
 \end{align}
 \end{subequations}
We omit rules that can be deduced by parity, for example $D_7* \bar N_3 = D_7 + N_3$. Modulo the interchiral and parity symmetries, these rules are dictated by null vectors, together with the constraint \eqref{777e} from the extended symmetry which leads to the appearance of only $D_7$ rather than $^eD_7$ in Eq. \eqref{d7d7}. Permutation symmetry does not add more constraints: for example, $C_{2,2,3}^\text{even}=0$ (dictated by null vectors) implies that $\left<V_{(2,s)}V_{(2,s)}V_{(3,s')}\right>=0$ if $\frac{p}{3}-s'$ is even, equivalently if $V_{(3,s')}$ has an odd conformal spin $3s'$ (as required by permutation symmetry).

As explained in Section \ref{sec:fr}, non-chiral fusion rules have no reason to be associative. For example,
\begin{subequations}
 \label{nonass}
\begin{align}
 D_4*(N_3*N_3) = D_4*(D_1+N_3) &= 2D_4+N_2 \ ,
 \\
 (D_4*N_3)*N_3=(D_4+N_2)*N_3 &= 2D_4+2N_2\ .
\end{align}
\end{subequations}
 
\subsection{Case $q=18$}

\begin{align}
\renewcommand{\arraystretch}{1.5}
 \begin{array}{|c|c|c|c|c|}
  \hline
  \text{Notation} & \text{Field} & \text{Values of } s & \text{Left} & \text{Right}
  \\
  \hline \hline
 D_1 &   V^d_{\langle 1,s\rangle} & ]0, p[ &  (1, s) & (1, s)
  \\
  \hline
  D_5 & V^d_{\langle 5,s\rangle} & ]0, p[ &  (5, s) & (5, s)
  \\
  \hline
  D_7 & V^d_{\langle 7,s\rangle} & ]0, p[ &  (7, s) & (7, s)
  \\
  \hline
  D_9 & V^d_{\langle 9,s\rangle} & ]0,p[ & (9, s) & (9, s)
  \\
  \hline \hline
  N_2 & V_{(2, s)} & ]-\frac{p}{2},\frac{p}{2}[ & (7,\frac{p}{2}-s) & (11,\frac{p}{2}-s)
  \\
  \hline
  N_4 & V_{(4, s)} & ]-\frac{p}{2},\frac{p}{2}[ & (5,\frac{p}{2}-s) & (13,\frac{p}{2}-s)
  \\
  \hline
  N_8 & V_{(8, s)} & ]-\frac{p}{2},\frac{p}{2}[ & (1,\frac{p}{2}-s) & (17,\frac{p}{2}-s)
  \\
  \hline
  N_3 & V_{(3,s)} &  ]-\frac{2p}{3},\frac{p}{3}[ & (3, \frac{p}{3}-s)  & (9,\frac{p}{3}-s)
  \\
  \hline
  \bar N_3 & V_{(3,s)} &  ]-\frac{p}{3},\frac{2p}{3}[ & (9, \frac{p}{3}+s)  & (3,\frac{p}{3}+s)
  \\
  \hline
 \end{array}
\end{align}
The distinction between even and odd couplings arises from solving the ambiguity $\mathcal{R}_{r,s}=\mathcal{R}_{q-r,p-s}$ by choosing a value in $\{r,q-r\}$, but this is not possible for $D_9$ as $r=q-r$ in this case. As a result, whenever $D_9$ is involved, there is no distinction between odd and even couplings.

The simple current $N_8$ generates a $\mathbb{Z}_2$ symmetry with the orbits 
\begin{align}
 \{D_9\}, \{N_3\},\{\bar N_3\}, \{D_1,N_8\},\{D_5, N_4\}, \{D_7, N_2\}\ . 
 \label{n8}
\end{align}
This means that we have the fusion rules 
\begin{align}
 N_8* D_9 = D_9 \quad , \quad N_8* D_5 = N_4 \quad , \quad N_8* N_4 = D_5 \quad , \quad \text{etc}.
\end{align}
Modulo this simple current symmetry, and modulo parity, the non-chiral fusion rules are 
\begin{subequations}\label{18fus}
\begin{align}
D_1* X &= X \quad (\forall X)\ ,
\\
D_5 * D_5 &= D_1 +  D_5 + D_7 + D_9 + N_3 + \bar N_3 \ ,
\\
D_5 * D_7 &= D_5 + {}^eD_7 +D_9 +{}^e N_2 + N_3 + \bar N_3 \ ,
\\
D_5 * D_9 &= D_5 + D_7 +D_9 +N_2 + N_4 \ ,
\\
D_5 * N_3 &=  D_5 +D_7 +N_2 + N_3 +N_4 \ ,
\\
D_7 * D_7 &= D_1 +  {}^eD_5 + {}^eD_7 + D_9 + N_3+  \bar N_3  +{}^eN_4 \ ,
\\
D_7 * D_9 &= D_5 + D_7 + N_2 + N_3 + \bar N_3 +N_4 \ ,
\\
D_7 * N_3 &= D_5 + D_7+ D_9 + N_2 +{}^e\bar N_3  +N_4 \ ,
\\
D_9 * D_9 &= D_1 +  D_5  + D_9  +N_4 + N_8 \ ,
\\
D_9 * N_3 &=  D_7   +N_2 + \bar N_3 \ ,
\\
N_3 * N_3 &= D_1 + D_5 + N_3 + N_4 + N_8 \ ,
\\
 N_3 * \bar N_3 & = {}^e D_7 + D_9 + {}^eN_2 \ .
\ . 
\end{align}
\end{subequations}
To be complete, let us write the remainder of the fusion rules, deduced using the relation
\eqref{mijk} applied to the simple current symmetry \eqref{n8}:
\begin{subequations}
\begin{align}
D_5 * N_2 &=  {}^e D_7 +D_9 +{}^eN_2 + N_3  +  \bar N_3 +N_4
\ , \\ 
D_5 * N_4 &=  D_9 +N_2 + N_3 + \bar N_3 +N_4 +N_8
\ , \\ 
D_7 * N_2 &= {}^eD_5 + D_9 + {}^eN_2 + N_3 + \bar N_3  +{}^eN_4 +N_8
\ , \\ 
D_7 * N_4 &= {}^eD_7+ D_9 + {}^eN_2 + N_3 +\bar N_3 +N_4
\ , \\ 
D_9 * N_2 &= D_5  + D_7  +N_2 + N_3 + \bar{N}_3 + N_4
\ , \\ 
D_9 * N_4 &= D_5 + D_7 + D_9   +N_2 + N_4
\ , \\ 
N_2 * N_2 &= D_1 + {}^eD_5  + {}^eD_7 + D_9 + N_3  +  \bar N_3+ {}^eN_4
\ , \\ 
N_2 * N_3 &=  D_5  + D_7 + D_9 + N_2+  {}^e\bar N_3 + N_4
\ , \\ 
N_2 * N_4 &=  D_5  + {}^eD_7 + D_9 + {}^eN_2 + N_3 +  \bar N_3
\ , \\
N_3 * N_4 &= D_5 + D_7 + N_2 + N_3 + N_4
\ , \\
N_4 * N_4 &= D_1 + D_5 + D_7 +D_9 + N_3 +\bar N_3 \ .
\end{align}
\end{subequations}
In these results, the vanishings $\left<D_7D_7N_2\right>=\left<D_9D_9N_2\right>=0$ are due to permutation symmetry. From $\left<D_9D_9N_2\right>=0$ we deduce $\left<D_9D_9D_7\right>=0$ using simple current symmetry. 
Similarly, $\left<V^d_{\langle 9,s_1\rangle}V^d_{\langle 9,s_1\rangle}N_{(3,s_3)}\right>$ vanishes by permutation symmetry if the spin of $N_{(3,s_3)}$ is odd. However, this coupling involves $D_9$, so there is no distinction between even and odd couplings. This implies that the couplings $\left<V^d_{\langle 9,s_1\rangle}V^d_{\langle 9,s_2\rangle}N_{(3,s_3)}\right>$ for all values of $s_i$ are related by interchiral symmetry, therefore $\left<D_9D_9N_3\right>=0$. We conclude that the non-chiral fusion rules are completely determined by the constraints of Section \ref{sec:con}.

\subsection{Case $q=30$}

\begin{align}
\renewcommand{\arraystretch}{1.5}
 \begin{array}{|c|c|c|c|c|}
  \hline 
  \text{Notation} & \text{Field} & \text{Values of } s & \text{Left} & \text{Right} 
  \\
  \hline \hline 
 D_1 & V^d_{\langle 1,s\rangle} & ]0, p[ &  (1, s) & (1, s) 
  \\
  \hline 
  D_7  & V^d_{\langle 7,s\rangle} & ]0, p[ &  (7, s) & (7, s) 
  \\
  \hline 
  D_{11}  & V^d_{\langle 11,s\rangle} & ]0, p[ &  (11, s) & (11, s) 
  \\
  \hline
  D_{13}   & V^d_{\langle 13,s\rangle} & ]0,p[ & (13, s) & (13, s)
  \\
  \hline \hline 
  N_2  & V_{(2, s)} & ]-\frac{p}{2},\frac{p}{2}[ & (13,\frac{p}{2}-s) & (17,\frac{p}{2}-s) 
  \\
  \hline 
  N_4 & V_{(4, s)} & ]-\frac{p}{2},\frac{p}{2}[ & (11,\frac{p}{2}-s) & (19,\frac{p}{2}-s) 
  \\
  \hline 
  N_8  & V_{(8, s)} & ]-\frac{p}{2},\frac{p}{2}[ & (7,\frac{p}{2}-s) & (23,\frac{p}{2}-s) 
   \\
  \hline 
  N_{14} & V_{(14, s)} & ]-\frac{p}{2},\frac{p}{2}[ & (1,\frac{p}{2}-s) & (29,\frac{p}{2}-s) 
  \\
  \hline 
  N_3  & V_{(3,s)} &  ]-\frac{2p}{3},\frac{p}{3}[ & (7, \frac{p}{3}-s)  & (13,\frac{p}{3}-s) 
  \\
  \hline 
  \bar N_3  & V_{(3,s)} &  ]-\frac{p}{3},\frac{2p}{3}[ & (13, \frac{p}{3}+s)  & (7,\frac{p}{3}+s) 
  \\
  \hline 
   N_9 & V_{(9,s)} &  ]-\frac{2p}{3},\frac{p}{3}[ & (1, \frac{p}{3}-s)  & (19,\frac{p}{3}-s) 
  \\
  \hline 
  \bar N_9 & V_{(9,s)} &  ]-\frac{p}{3},\frac{2p}{3}[ & (19, \frac{p}{3}+s)  & (1,\frac{p}{3}+s) 
  \\
  \hline 
  N_5  & V_{(5,s)} & ]-\frac{4p}{5}, \frac{p}{5}[ &  (1,\frac{p}{5}-s) & (11, \frac{p}{5}-s) 
  \\
  \hline 
  \bar N_5    & V_{(5,s)} & ]-\frac{p}{5}, \frac{4p}{5} [ & (11,\frac{p}{5}+s) & (1,\frac{p}{5}+s)
  \\
  \hline 
  N_5'  & V_{(5,s)} & ]-\frac{3p}{5},\frac{2p}{5}[ & (7,\frac{2p}{5}-s) & (17,\frac{2p}{5}-s) 
  \\
  \hline 
  \bar N_5'  & V_{(5,s)} & ]-\frac{2p}{5},\frac{3p}{5}[ & (17,\frac{2p}{5}+s) & (7,\frac{2p}{5}+s) 
  \\
  \hline 
 \end{array}
 \label{spec30}
\end{align}
The simple current $N_{14}$ generates a $\mathbb{Z}_2$ symmetry whose orbits all have length $2$:
\begin{align}
 \{D_1, N_{14}\} , \{D_7,N_8\} , \{D_{11},N_4\} , \{D_{13},N_2\}, \{N_3,N_5'\}, \{N_9,N_5\}\ .
 \label{n14}
\end{align}
Modulo this simple current symmetry, and modulo, parity, the fusion rules are 
\begin{subequations}
\begin{align}
D_7 * D_7 &= D_1 +D_7 + D_{11} + D_{13} + N_3 + \bar N_3 + N_5 + \bar N_5
\ ,  \\
D_7 * D_{11} &= {}^eD_{13} + {}^eN_2 + N_3 + \bar N_3  + N_5' + \bar N_5'
\ ,\\
D_{7} * D_{13} &= D_{7} + {}^eD_{11} +{}^eD_{13}+ {}^eN_2 + N_3 + \bar N_3 +{}^eN_4 +N_5' + \bar N_5'
\ , \\
D_{7} *  N_3 &=
D_7 + D_{11} + D_{13} +N_2 + N_3 + \bar N_3 + N_4 + N_5 +  N_5' + N_9
\ , \\
D_7 * N_9
& =
 N_3 +  N_5' + N_8
\ , \\
D_{11} * D_{11}
&= D_1 + {}^e N_4  +\bar N_5 + N_5 + \bar N_9 + N_9
\ , \\
D_{11} * D_{13}
&= {}^eD_7  + {}^eD_{13} + {}^eN_2 + {}^e\bar N_3 + {}^eN_3  + {}^e\bar N_5' + {}^eN_5' + N_8
\ , \\
D_{11} * N_3
&= D_7 + {}^eD_{13} + {}^eN_2 + {}^e\bar N_3 + N_3  + {}^e\bar N_5' + N_5' + N_8
\ , \\
D_{11} * N_9
&= D_{11} + N_4 + \bar N_9
\ , \\
D_{13} * N_3
&= D_7 + {}^eD_{11} + D_{13} + {}^EN_2 + {}^e\bar N_3 + N_3 + {}^eN_4 + {}^e\bar N_5' + \bar N_5 + N_8 + \bar N_9
\ , \\
D_{13} * N_9
&= D_{13} + N_2 + \bar N_3 + \bar N_5 '
\ , \\
N_3 * N_9
&= D_7 +  \bar N_3 + N_5' + N_8 
\ , \\
N_3 * N_3
&=
D_1 + D_7 + D_{11} + D_{13} + N_3 + \bar N_3 +N_4  + N_5 + \bar N_5 + \bar N_5'  + N_8 + N_9
\ , \\
N_9 * N_9
& =
D_1 + N_5 + N_9
\ , \\
D_{13} * D_{13}
&=
D_{1} + {}^eD_7 +{}^eD_{11} + D_{13} + N_3 + \bar N_3 + {}^eN_4  
\nonumber
\\
&\hspace{3cm}+ N_5 + \bar N_5 + {}^EN_5' + {}^e\bar N_5'+ N_8 + N_9 + \bar N_9 
\ , \\
N_3 * \bar N_3 
&= D_7 + {}^eD_{11} + {}^eD_{13}  + {}^eN_2 + N_3 + \bar N_3 + {}^eN_4 + N_5' + \bar N_5'\ ,
\end{align}
\end{subequations}
where we have 2 instances of the notation $^E$ which indicates that only even couplings appear, whereas the more common notation $^e$ means that both even and odd couplings appear.

In particular, some terms are absent due to the extended symmetry \eqref{1111x}, namely
\begin{align}
 C_{11,11,2}= C_{11,11,13}= C_{11,11,3}  = C_{11,11,5'} = C_{11,11,7}= C_{11,11,8}  = 0\ .
 \label{c0}
\end{align}
To get the full consequences of the extended symmetry, we should act on these couplings as in Eq. \eqref{mijk}, using the simple current $\mathbb{Z}_2$ symmetry \eqref{n14}. 

Permutation invariance implies $C_{X,X,2}=0$. Then, by $\mathbb{Z}_2$ symmetry,  $C_{3,3,2}=0 \implies C_{13,3,5'}=0$. Furthermore, using the relation between integer and fractional Kac indices in Table \eqref{spec30}, we can translate the permutation invariance condition \eqref{cxxy} into a condition on odd or even couplings, and we find the constraints
\begin{align}
 C_{X,X,9}^\text{even}=C_{X,X,5}^\text{even}=C_{X,X,3}^\text{even}=C_{X,X,5'}^\text{odd}=0\ .
\end{align}
The first two constraints already follow from null vectors, but the last two lead to the nontrivial vanishings
\begin{align}
C_{13,13,3}^\text{even} = C_{13,2,5'}^\text{even} = C_{2,2,3}^\text{even} = C_{13,13,5'}^\text{odd}=C_{13,2,3}^\text{odd} = C_{2,2,5'}^\text{odd}=0\ . 
\end{align}

There remains one coupling whose vanishing is a purely numerical finding, without any motivation from the constraints we have discussed:
\begin{align}
 \boxed{C^\text{even}_{13,13,13}=0}\ .
\end{align}
This is formally similar to the vanishing \eqref{777e} for the $E_{p,12}$ models, which also involves the diagonal field with the largest first Kac index. However, in the $E_{p,12}$ case, the vanishing was explained by the extended symmetry. As this is our only unexplained vanishing, we have checked that it occurs in several 4-point functions. In the case $p=7$, we have bootstrapped $\left<V^d_{\langle 13,1\rangle}V^d_{\langle 13,1\rangle}V^d_{\langle 13,1\rangle}V^d_{\langle 13,1\rangle}\right>$, $\left<V^d_{\langle 11,1\rangle}V^d_{\langle 13,1\rangle}V^d_{\langle 13,1\rangle}V^d_{\langle 13,1\rangle}\right>$ and $\left<V^d_{\langle 13,1\rangle}V^d_{\langle 13,1\rangle}V^d_{\langle 11,1\rangle}V^d_{\langle 3,\frac43\rangle}\right>$. As we increase the numerical precision, we find that the 4-point structure constant $D^{(s)}_{(13,1)}$ (which includes a factor $C^\text{odd}_{13,13,13}$) converges to a finite value, while $D^{(s)}_{(13,6)}$ (which includes a factor $C^\text{even}_{13,13,13}$) tends to zero. We have observed these tendencies up to about $38$ significant digits.

\section{Structure constants}\label{sec:sc}

\subsection{General form}

Assuming the spectrums \eqref{specs} and the chiral fusion rules, we numerically find a unique solution of the crossing symmetry equation for any given 4-point function. This solution gives us access to the 4-point structure constants \eqref{4pt}. From there, we can in principle determine 3-point structure constants modulo field renormalizations $V_i\to \lambda_i V_i$, but it is challenging to extract analytic formulas. To do this, we will first divide the numerical results by a reference structure constant, the same as in loop models \cite{nrj23}. After that, again as in loop models, we will asssume that we obtain simple functions of
\begin{align}
 n = -2\cos\left(\pi \frac{p}{q}\right) \ . 
 \label{npq}
\end{align}
This assumption is well-motivated in the case of loop models, where $n$ is interpreted as the statistical loop weight. We do not have such a motivation in the case of E-series minimal models, although their lattice description \cite{pas87} might help if it could be generalized to all values of $p$. Nevertheless, the assumption
will turn out to be obeyed by all our numerical results. 

Let us first write the reference 3-point structure constant:
\begin{align}
C^\text{ref}_{(r_1,s_1)(r_2,s_2)(r_3,s_3)} =\prod_{\epsilon_1,\epsilon_2,\epsilon_3=\pm} \Gamma_\beta^{-1} \left(\tfrac{\beta+\beta^{-1}}{2} + \tfrac{\beta}{2}\left|\textstyle{\sum_i} \epsilon_ir_i\right| + \tfrac{\beta^{-1}}{2}\textstyle{\sum_i} \epsilon_is_i\right)\ , 
 \label{cref}
\end{align}
where $\beta = \sqrt{\frac{p}{q}}$, and 
the term $\left|\textstyle{\sum_i} \epsilon_ir_i\right|$ involves an absolute value. This ansatz is valid for 3-point functions of non-diagonal fields in the notation $V_{(r,s)}$ \eqref{vrs}. It is also valid if diagonal fields are present, provided we write a diagonal field with left and right dimensions $\Delta=\bar\Delta=\frac{c-1}{24}+P^2$ using the non-diagonal notation $V_{(0,2\beta P)}$. We then deduce reference 2-point and 4-point structure constants via Eqs. \eqref{bc} and \eqref{4pt}. After dividing structure constants by their reference versions, we obtain the \textbf{reduced structure constants}
\begin{align}
 b_i = \frac{B_i}{B_i^\text{ref}} \quad , \quad c_{i,j,k} = \frac{C_{i,j,k}}{C_{i,j,k}^\text{ref}} \quad , \quad d_k^{(s)} = \frac{D_k^{(s)}}{D_k^{(s),\text{ref}}}\ . 
 \label{genc}
\end{align}
By construction, reference structure constants obey the constraints from interchiral symmetry modulo signs. As a result, the ratios $c_{(r_1,s_1)(r_2,s_2)(r_3,s_3)}$ only depend on the second Kac indices $s_i$ through overall signs, and these signs are determined by
 \begin{align}
 \frac{c_{(r_1,s_1+1)(r_2,s_2)(r_3,s_3)}}{c_{(r_1,s_1-1)(r_2,s_2)(r_3,s_3)}} &\ =\  (-1)^{\max(2r_1, 2r_2, 2r_3,r_1+r_2+r_3)} \ ,
 \label{tpt}
 \\
 \frac{c_{(r_1,s_1+1)(r_2,s_2)(r_3,s_3)}}{c_{(r_1,s_1)(r_2,s_2)(r_3,s_3+1)}} &= 
 \left\{\begin{array}{ll} (-1)^{r_1+r_2} &\text{ if } r_2\geq |r_1-r_3|\ ,
                        \\ (-1)^{r_3} &\text{ else}\ .
                       \end{array}\right. 
\label{tppt}
\end{align}

Let us now discuss the dependence on $n$ \eqref{npq}. For $p \in \mathbb{N}$ such that $p$ and $q$ are coprime, $n$ takes finitely many values, which are the roots of an even polynomial $P_q$:
\begin{align}
\renewcommand{\arraystretch}{1.3}
 \begin{array}{|c|l|}
  \hline 
  q &  P_q(n) 
  \\
  \hline 
  12 & n^4-4n^2+1 
  \\
  18 & n^6 -6n^4 +9n^2-3 
  \\
  30 & n^8 - 7n^6+14n^4-8n^2+1 
  \\
  \hline 
 \end{array}
\end{align}
Two of these polynomials have simple expressions in terms of Chebyshev polynomials of the first kind, namely $P_{12}(n) = 2T_4(\tfrac{n}{2})-1$ and $P_{18}(n) = 2T_6(\tfrac{n}{2})-1$. Any rational function of $n$ may be rewritten as a polynomial of $n$ of degree $\deg P_q -1$. The analogy with loop models leads us to the conjecture: 
\begin{quote}
 \textbf{Conjecture:} Reduced structure constants are polynomial functions of $n$ with values in the field $\mathbb{Q}(\cos(\frac{\pi}{q}))$.
\end{quote}
This does not mean that reduced structure constants are polynomial with rational coefficients, i.e. elements of $\mathbb{Q}[n]$. For example, in the case $q=12$, we have $n(n^2-3)=\pm \sqrt{2}$, with a $p$-dependent sign. The constant $\sqrt{2}$ is a polynomial function of $n$ with an irrational coefficient, and for all allowed $n$ its value belongs to the field $\mathbb{Q}(n)= \mathbb{Q}(\cos(\frac{\pi}{12}))$, but it cannot be rewritten as a polynomial with rational coefficients. 

\subsection{Explicit expressions for $q=12$}

Let us write the $4$ values of $n$ in this case. We also introduce signs $\mu,\nu\in\{-1,1\}$, which will be convenient for writing structure constants:
\begin{align}
\renewcommand{\arraystretch}{1.4}
 \begin{array}{|c||c|c|c|c|}
 \hline 
  p\bmod 24 & 1,23 & 5,19 & 7,17 & 11,13 
  \\
  \hline
  n & -\frac{\sqrt{3}+1}{\sqrt{2}} & -\frac{\sqrt{3}-1}{\sqrt{2}} & \frac{\sqrt{3}-1}{\sqrt{2}} & \frac{\sqrt{3}+1}{\sqrt{2}}
  \\
  \hline 
  n^2 & 2+\sqrt{3} & 2-\sqrt{3} & 2-\sqrt{3} & 2+\sqrt{3} 
  \\
  \hline
  \mu & 1 & -1 & 1 & -1
  \\
  \hline 
  \nu & 1 & 1 & -1 & -1
  \\
  \hline 
 \end{array}
\end{align}
Equivalently, we have 
\begin{align}
 n = -\frac{\mu+ \sqrt{3} \nu}{\sqrt{2}} \quad , \quad n^2 = 2+\sqrt{3}\mu\nu \quad ,\quad n^{-1} = \frac{\mu-\sqrt{3}\nu}{\sqrt{2}}\ .
\end{align}
Let us now write our results for 2-point and 3-point structure constants, where:
\begin{itemize}
 \item We normalize fields such that $b_1=b_4=b_7=1$, but we find it convenient to have nontrivial $b_2$ and $b_3=b_{\bar 3}$. We include these 2-point structure constants in the following table under the names $c_{1,2,2}$ and $c_{1,3,3}$.
 \item For any 3-point structure constant $c_{r_1,r_2,r_3}$ we call $s_1,s_2,s_3$ the second Kac indices, whose values are displayed in Table \eqref{spec12}. 
 \item We display the sign that is picked under odd permutations of the 3 fields, which is $-1$ if the sum of conformal spins is odd. 
 \item We display the condition on the parity of $\sum s_i =s_1+s_2+s_3$, under which each structure constant is valid. If all 3 fields are diagonal, this is simply $\sum s_i\in 2\mathbb{Z}+1$ in the odd case and $\sum s_i+1\in 2\mathbb{Z}+1$ in the even case. If non-diagonal fields are involved, this can be more complicated due to the map between integer and fractional Kac indices, see Table \eqref{spec12}. We write superscripts ${}^\text{even}$ in even cases, and no superscript in odd cases.
 \item To obtain compact formulas, we find it convenient to write structure constants as powers of $n$ up to signs and constant prefactors, although in view of the polynomial equation $P_{12}(n)=0$ there are many equivalent expressions.
\end{itemize}

\begin{align}
\renewcommand{\arraystretch}{1.5}
 \begin{array}{|c|r|l|r|}
  \hline 
  \text{Name} & \text{Sign} & \in 2\mathbb{Z}+1 & \text{Value}
  \\
  \hline \hline 
   c_{1,2,2} & 1 & \sum s_i + 1 & -1
   \\
   \hline 
   c_{1,3,3} & (-1)^{3s_2+3s_3}& \sum s_i + \frac{2p}{3} &  -(-1)^{3s_3}\nu\sqrt{6}
   \\
    \hline \hline 
   c_{2,4,7} & -1& \sum s_i -\frac{p}{2} & -\sqrt{\frac32} \cdot n^{-1}
   \\
  \hline 
   c_{2,4,7}^\text{even} & -1& \sum s_i + \frac{p}{2} & \sqrt{\frac32} \cdot n^{-1}
   \\
  \hline 
  c_{4,4,7} & 1& \sum s_i & \nu\sqrt{\frac12}
  \\
  \hline 
   c_{4,4,7}^\text{even} & 1& \sum s_i +1 & \mu\sqrt{\frac32}
   \\
   \hline
   c_{2,2,7} & 1& \sum s_i +1 & -\mu\nu\frac12 \cdot n^{-3}
   \\
   \hline 
    c_{2,2,7}^\text{even} & 1& \sum s_i & -\sqrt{\frac34} \cdot n^{-1}
    \\
    \hline 
    c_{7,7,7} & 1& \sum s_i& \nu\sqrt{2}
    \\
    \hline 
    c_{4,4,3} & (-1)^{3s_3}& \sum s_i -\frac{p}{3} & (-1)^{s_1}\sqrt{2}\cdot n^{-1}
    \\
    \hline 
    c_{7,7,3} & (-1)^{3s_3}& \sum s_i-\frac{p}{3} & (-1)^{s_1}\nu\sqrt{12}\cdot n^{-1}
    \\
    \hline 
    c_{7,3,\bar 3} & (-1)^{3s_2+3s_3}& \sum s_i & (-1)^{3s_3}\mu\sqrt{6}\cdot n^{-1}
    \\
    \hline 
    c_{3,3,3} & 1 & \sum s_i+1 & (-1)^{\frac12(s_1+s_2+s_3)} \mu\nu\sqrt{6}\cdot n^{-2} 
    \\
    \hline 
    c_{2,2,3} & (-1)^{3s_3} & \sum s_i +\frac{2p}{3} &(-1)^{\frac12(s_1-s_2-3s_3)} \sqrt{\frac12} \cdot n^{-3}
    \\
    \hline 
     c_{2,4,3} & -(-1)^{3s_3} & \sum s_i -\frac{5p}{6} &-(-1)^{s_1+\frac{p}{2}} \mu\nu\sqrt{3}\cdot n^{-1}
    \\
    \hline 
 \end{array}
 \label{12tab}
\end{align}
Apart from the structure constants $c_{1,2,2}$ and $c_{1,3,3}$, these are the $13$ nonzero independent structure constants of the $E_{p,12}$ minimal models. After extracting these structure constants, we have tested them by directly checking crossing symmetry of a variety of 4-point functions \cite{rng24}.

\subsection{Tests of the conjecture for $q=18,30$}

The large spectrums of the $E_{p,18}$ and $E_{p,30}$ models lead to two difficulties when numerically computing structure constants:
\begin{enumerate}
 \item There are many structure constants to be determined. 
 \item The decompositions of a given 4-point function into conformal blocks can involve many terms. 
\end{enumerate}
We will address these difficulties by testing the conjecture in a few 4-point functions only, and choosing 4-point functions that involve not too many different structure constants. These include 4-point functions where at least one of the 4 fields is a simple current. These also include 4-point functions where some or all 4 fields are identical, so that some or all of the 3 channels $s,t,u$ involve the same structure constants.

We numerically determine 4-point structure constants using the same methods as in loop models \cite{rib24}. We then find analytic formulas under the assumption that the conjecture holds, i.e. that reduced structure constants are polynomial functions of $n$ \eqref{npq}. To find these polynomials, we compute the structure constants for a large enough number of values of $n$, equivalently of $p$. 
In practice we find it convenient to write our results as rational functions of $n$, rather than polynomials. Once we have an analytic formula, we compute the error, defined as
\begin{equation}
\text{Error}
= \left|1-\frac{\text{numerical value}}{\text{analytic formula}}\right|\ .
\end{equation}
We find that this error is always of the same order of magnitude as the deviation, which estimates the numerical uncertainty of our bootstrap method \cite{rib24}. In particular, the error decreases when we push the numerics to better precision. In the results that we display, the error is computed in the case $p=17$.

We will now display the comparison between the analytic structure constants and numerical results from the bootstrap, in the cases of structure constants with smallest second indices.

\paragraph{$\big\langle V_{\langle 5, 1 \rangle}^dV_{\langle 5, 1 \rangle}^dV_{\langle 5, 1 \rangle}^dV_{\langle 5, 1 \rangle}^d \big\rangle$ for $q = 18$}

\begin{align}
\renewcommand{\arraystretch}{1.3}
 \begin{array}{|c|c|c|}
  \hline 
  \text{Structure constant} & \text{Value} &  \text{Error} 
  \\
  \hline \hline 
  d_{\langle 1,1 \rangle^d}^{(s,t,u)} & 1 & 0
  \\
  \hline 
  d_{\langle 5,1 \rangle^d}^{(s,t,u)} & 1  &  10^{-40}
  \\
  \hline 
   d_{\langle 7,1 \rangle^d}^{(s,t,u)} & 1  & 10^{-40}
  \\
  \hline 
   d_{\langle 9,1 \rangle^d}^{(s,t,u)} & \frac12  & 10^{-37}
  \\
  \hline 
   d_{(3,\frac13)}^{(s,t,u)} & \frac{6}{n^3}  & 10^{-16}
  \\
  \hline 
 \end{array}
\end{align}

\paragraph{$\big\langle V_{\langle 5, 1 \rangle}^dV_{( 8, \frac12)}V_{\langle 5, 1 \rangle}^dV_{( 8, \frac12)} \big\rangle$ for $q = 18$}

\begin{align}
\renewcommand{\arraystretch}{1.3}
 \begin{array}{|c|c|c|}
  \hline 
  \text{Structure constant} & \text{Value} &   \text{Error} 
  \\
  \hline \hline 
d_{\langle 1,1 \rangle^d}^{(u)} & 1  & 0
  \\
  \hline 
d_{( 4,\frac12)}^{(s,t)} &   18\cdot\frac{8-3n^2}{5 - 11n^2}
 &10^{-29}
  \\
  \hline 
 \end{array}
\end{align}

\paragraph{$\big\langle V_{(3, \frac23)}V_{(3, \frac23)}V_{(3, \frac23)}V_{(3, \frac23)} \big\rangle$ for $q = 18$}

\begin{align}
\renewcommand{\arraystretch}{1.3}
 \begin{array}{|c|c|c|}
  \hline 
  \text{Structure constant} & \text{Value}  & \text{Error} 
  \\
  \hline \hline 
d_{\langle 1,s \rangle^d}^{(s,t,u)} & 1&  0 
  \\
  \hline 
d_{(3, \frac23)}^{(s,t,u)}&   
 -2n(n^2 - 4)
   &
10^{-37}
  \\
  \hline 
d_{(4, -\frac12)}^{(s,t,u)}
&   
3\cdot\frac{n^2 - 3}{n^2 - 1} 
&
 10^{-21}
  \\
  \hline 
d_{\langle 5,1 \rangle^d}^{(s,t,u)} &
 -n^2 + 4
  &
  10^{-39}
  \\
  \hline 
  d_{( 8, - \frac12)}^{(s,t,u)} &
   9\cdot\frac{2n^2 - 9}{4 - 3n^2}
      &
 10^{-13}
  \\
  \hline 
 \end{array}
\end{align}

\paragraph{$\big\langle V_{\langle 7, 1 \rangle}^dV_{\langle 7, 1 \rangle}^dV_{\langle 7, 1 \rangle}^dV_{\langle 7, 1 \rangle}^d \big\rangle$ for $q = 30$}

\begin{align}
\renewcommand{\arraystretch}{1.3}
 \begin{array}{|c|c|c|}
  \hline 
  \text{Structure constant} & \text{Value} &   \text{Error} 
  \\
  \hline \hline 
d_{\langle 1,1 \rangle^d}^{(s,t,u)} & 1 &  0 
  \\
  \hline 
d_{(3, \frac{14}{3})}^{(s,t,u)}&   
-10\cdot\frac{2n^4 -10n^2 + 7}{9n^3}
   &
  10^{-20}
  \\
  \hline 
d_{(5, \frac{12}{5})}^{(s,t,u)}
&   
-62 n^7+428 n^5
-828 n^3+416 n
&
10^{-17}
  \\
  \hline 
d_{\langle 7,1 \rangle^d}^{(s,t,u)} &
\frac53
  &
  10^{-21}
  \\
  \hline 
d_{\langle 11,1 \rangle^d}^{(s,t,u)} &
1
  &
  10^{-18}
  \\
  \hline 
d_{\langle 13,1 \rangle^d}^{(s,t,u)} &
\frac13
  &
  10^{-15}
  \\
  \hline 
 \end{array}
\end{align}

\paragraph{$\big\langle V_{\langle 7, 1 \rangle}^dV_{( 14, \frac12)}V_{\langle 7, 1 \rangle}^dV_{( 14, \frac12)} \big\rangle$ for $q = 30$}

\begin{align}
\renewcommand{\arraystretch}{1.3}
 \begin{array}{|c|c|c|}
  \hline 
  \text{Structure constant} & \text{Value} &   \text{Error} 
  \\
  \hline \hline 
d_{\langle 1,1 \rangle^d}^{(u)} & 1 &  0 
  \\
  \hline 
d_{(8, \frac12)}^{(s,t)}&   
-30 \big(57 n^6-389 n^4
+730 n^2-329\big)
   &
  10^{-48}
  \\
  \hline 
 \end{array}
\end{align}

\section*{Acknowledgements}

We are grateful to Ingo Runkel, Vincent Pasquier, Valentin Benedetti, Yoshiki Fukusumi, and Ken Kikuchi, for enlightening discussions and correspondence. Moreover, we wish to thank Vincent Pasquier for helpful suggestions on the draft article, and Jiaxin Qiao for pointing out a typo in Table \eqref{12tab}.

We wish to thank Connor Behan and Jiaxin Qiao, who reviewed this article for SciPost, for their questions and suggestions, which led to many clarifications in this text.

This work is partly a result of
the project ReNewQuantum, which received funding from the European Research Council. 

\bibliographystyle{morder7}
\bibliography{992}

@article{fgp90,
    author = "Furlan, P. and Ganchev, A. Ch. and Petkova, V. B.",
    title = "{Fusion Matrices and $C < 1$ (Quasi)local Conformal Theories}",
    reportNumber = "INFN-AE-89-3",
    doi = "10.1142/S0217751X90001252",
    journal = "Int. J. Mod. Phys. A",
    volume = "5",
    pages = "2721--2736",
    year = "1990",
    note = "[Erratum: Int.J.Mod.Phys.A 5, 3641 (1990)]"
}

@misc{fuk24,
    author = "Fukusumi, Yoshiki",
    title = "{Fusion rule in conformal field theories and topological orders: A unified view of correspondence and (fractional) supersymmetry and their relation to topological holography}",
    eprint = "2405.05178",
    archivePrefix = "arXiv",
    primaryClass = "hep-th",
    month = "5",
    year = "2024"
}

@misc{mor20,
    author = "Moriwaki, Yuto",
    title = "{Full vertex algebra and bootstrap -- consistency of four point functions in 2d CFT}",
    eprint = "2006.15859",
    archivePrefix = "arXiv",
    primaryClass = "math.QA",
    month = "6",
    year = "2020"
}

@article{kp07,
   title={On Polynomials Interpolating Between the Stationary State of a O(n) Model and a Q.H.E. Ground State},
   volume={276},
   ISSN={1432-0916},
   eprint= "cond-mat/0608160",
   DOI={10.1007/s00220-007-0341-0},
   number={2},
   journal={Communications in Mathematical Physics},
   publisher={Springer Science and Business Media LLC},
   author={Kasatani, M. and Pasquier, V.},
   year={2007},
   month=sep, pages={397–435} }

@misc{rng24,
author = {{S. Ribault, R. Nivesvivat, L. Grans-Samuelsson et al}},
title = {Bootstrap\_Virasoro: Bootstrapping two-dimensional CFTs with Virasoro symmetry},
url = {https://gitlab.com/s.g.ribault/Bootstrap_Virasoro/},
year = {2024},
type = {code},
}

@misc{bcm24,
    author = "Benedetti, Valentin and Casini, Horacio and Magan, Javier M.",
    title = "{Selection rules for RG flows of minimal models}",
    eprint = "2412.16587",
    archivePrefix = "arXiv",
    primaryClass = "hep-th",
    month = "12",
    year = "2024"
}

@article{pas87,
title = {Two-dimensional critical systems labelled by Dynkin diagrams},
journal = {Nuclear Physics B},
volume = {285},
pages = {162-172},
year = {1987},
issn = {0550-3213},
doi = {https://doi.org/10.1016/0550-3213(87)90332-4},
url = {https://www.sciencedirect.com/science/article/pii/0550321387903324},
author = {Vincent Pasquier}
}

@misc{rs99,
    author = "Rida, A. and Sami, T.",
    title = "{Nonchiral fusion rules, structure constants of D(m) minimal models}",
    eprint = "hep-th/9910070",
    archivePrefix = "arXiv",
    reportNumber = "SUBATECH-00-04",
    month = "10",
    year = "1999"
}

@misc{rib24,
    author = "Ribault, Sylvain",
    title = "{Exactly solvable conformal field theories}",
    eprint = "2411.17262",
    archivePrefix = "arXiv",
    primaryClass = "hep-th",
    month = "11",
    year = "2024"
}

@article{cz09,
    author = "Cappelli, Andrea and Zuber, Jean-Bernard",
    title = "{A-D-E Classification of Conformal Field Theories}",
    eprint = "0911.3242",
    archivePrefix = "arXiv",
    primaryClass = "hep-th",
    doi = "10.4249/scholarpedia.10314",
    journal = "Scholarpedia",
    volume = "5",
    number = "4",
    pages = "10314",
    year = "2010"
}

@misc{rjrs24,
    author = "Roux, Paul and Jacobsen, Jesper Lykke and Ribault, Sylvain and Saleur, Hubert",
    title = "{Critical spin chains and loop models with $PSU(n)$ symmetry}",
    eprint = "2404.01935",
    archivePrefix = "arXiv",
    primaryClass = "hep-th",
    month = "4",
    year = "2024"
}

@misc{nrj23,
    author = "Nivesvivat, Rongvoram and Ribault, Sylvain and Jacobsen, Jesper Lykke",
    title = "{Critical loop models are exactly solvable}",
    eprint = "2311.17558",
    archivePrefix = "arXiv",
    primaryClass = "hep-th",
    month = "11",
    year = "2023"
}

@article{prz06,
    author = "Pearce, Paul A. and Rasmussen, Jorgen and Zuber, Jean-Bernard",
    title = "{Logarithmic minimal models}",
    eprint = "hep-th/0607232",
    archivePrefix = "arXiv",
    doi = "10.1088/1742-5468/2006/11/P11017",
    journal = "J. Stat. Mech.",
    volume = "0611",
    pages = "P11017",
    year = "2006"
}

@Article{NR20,
    author = "Nivesvivat, Rongvoram and Ribault, Sylvain",
    title = "{Logarithmic CFT at generic central charge: from Liouville theory to the $Q$-state Potts model}",
    eprint = "2007.04190",
    archivePrefix = "arXiv",
    primaryClass = "hep-th",
    doi = "10.21468/SciPostPhys.10.1.021",
    journal = "SciPost Phys.",
    volume = "10",
    number = "1",
    pages = "021",
    year = "2021"
}

@Book{fms97,
     author    = "Di Francesco, P. and Mathieu, P. and Sénéchal, D.",
     title     = "Conformal field theory",
     year = "1997",
     doi = "10.1007/978-1-4612-2256-9",
}

\end{document}